**Universal *in vivo* Textural Model for Human Skin based on Optical Coherence Tomograms**

Saba Adabi[1,2], Matin Hosseinzadeh[1], Shahryar Noei[1], Steven Daveluy[3,4], Anne Clayton[1], Darius Mehregan[4], Silvia Conforto[2], Mohammadreza Nasiriavanaki[1,3,4]

[1] Biomedical Engineering Department, Wayne State University, Detroit, Michigan, USA

[2] Applied Electronics Department, Roma Tre University, Via Volterra, Rome, Italy

[3] Department of Dermatology, Wayne State University School of Medicine, Detroit, Michigan, USA

[4] Barbara Ann Karmanos Cancer Institute, Detroit, Michigan, USA,

**Abstract**

Currently, diagnosis of skin diseases is based primarily on the visual pattern recognition skills and expertise of the physician observing the lesion. Even though dermatologists are trained to recognize patterns of morphology, it is still a subjective visual assessment. Tools for automated pattern recognition can provide objective information to support clinical decision-making. Noninvasive skin imaging techniques provide complementary information to the clinician. In recent years, optical coherence tomography (OCT) has become a powerful skin imaging technique. According to specific functional needs, skin architecture varies across different parts of the body, as do the textural characteristics in OCT images. There is, therefore, a critical need to systematically analyze OCT images from different body sites, to identify their significant qualitative and quantitative differences. Sixty-three optical and textural features extracted from OCT images of healthy and diseased skin are analyzed and, in conjunction with decision-theoretic approaches, used to create computational models of the diseases. We demonstrate that these models provide objective information to the clinician to assist in the diagnosis of abnormalities of cutaneous microstructure, and hence, aid in the determination of treatment. Specifically, we demonstrate the performance of this methodology on differentiating basal cell carcinoma (BCC) and squamous cell carcinoma (SCC) from healthy tissue.

**Introduction**

Visual inspection of skin abnormalities is the first step in diagnosis; however, it has limitations. Experienced dermatologists and surgeons are able to reach a diagnostic accuracy of about 75% with visual inspection only[1]. In order to confirm diagnosis, a biopsy may be performed. While biopsy is currently the gold standard diagnostic method for suspicious skin lesions, it may be inconvenient for the patient and create scarring. In addition, there are costs to the healthcare system and, potentially, increased morbidity related to the procedure. To assist dermatologists and improve diagnostic accuracy, ancillary methods of diagnosis have been sought including dermoscopy[2], reflectance confocal microscopy (RCM)[3], high frequency ultrasound (HFUS)[4], Magnetic Resonance Imaging (MRI)[5], diffuse multispectral imaging (MS)[6], and Raman spectroscopy[7]. More recently optical coherence tomography (OCT) has been introduced as an effective high-resolution cutaneous imaging technique with a moderate penetration depth for non-invasive inspection of the skin[8]. OCT allows *in situ*, safe, real-time investigation of micromorphology and pathology without tissue removal. To form an OCT image, the magnitude and time delay of backscattered light returned from a biological sample is measured transversally[9]. The resolution, ranging from about 1 μm in ultra-high-resolution systems to around 20 μm, is superior to that achieved by other tomographic methods such as conventional ultrasound, enabling cell-level detail to be observed[10]. OCT is therefore of interest for both *in vitro* and *in vivo* investigations of epithelial tissues. Its ability to perform 'optical biopsy' gives it the potential to replace, or at least to reduce the necessity for invasive tissue sampling via biopsy[11]. OCT has been studied to evaluate a variety of skin disorders including tumors, various inflammatory and blistering conditions, physical and chemical skin damage, as well as surgical interventions[12-17].

OCT images visualize the morphological details of tissue microstructures, i.e., stratum corneum, epidermis, dermis, hair follicles, eccrine sweat ducts, and sebaceous glands[9,10,18]. Figure 1 illustrates some of the structures in healthy skin visible in OCT images. The basic structure of a healthy skin includes the epidermis, dermis, and subcutaneous fat. The epidermis is four to five layers of stratified epithelia with no blood vessels, the most superficial being the stratum corneum[19]. The epidermis connects to the dermis by a layer known as the dermal-epidermal junction (DEJ). Cutaneous appendages, including sensory receptors, nerves, glands, blood vessels and hair follicles, reside in the dermis. Skin varies in color, thickness, and texture in different parts of the body according to specific functional needs. Regional variations include thickness of the stratum corneum, the presence of a stratum lucidum on palms and soles, epidermal thickness and variable numbers of sebaceous glands, eccrine glands and hair follicles[19]. In this study, we have looked at nose, preauricular, neck, volar forearm, palm, back, thumb, dorsal forearm, sole, and calf as representative of the variety of skin architectures and epidermal thicknesses across the body. The most notable features of thick skin (palm, thumb and sole) are the thick stratum corneum, presence of a stratum lucidum, an abundance of eccrine sweat glands and, a lack of hair follicles, sebaceous glands and apocrine glands. In OCT images of skin from the palm and sole, the stratum corneum is the first visualized layer of the epidermis, appearing as a homogenous layer of cells with scattered eccrine sweat ducts. The eccrine sweat ducts of thick skin have a recognizable spiral lumen when observed with high intensity reflected light, a result of the large refractive index mismatch between sweat duct and the keratinocytes of the epidermis[20]. The stratum lucidum, a clear thin layer of dead cells found only on the thickened epidermis of palms and soles, is just beneath the stratum corneum[10]. The prominent morphological features of the skin of the nose, preauricular, volar forearm, neck, back, dorsal forearm and calf are: thinner epidermis, no stratum lucidium and presence of hair and sebaceous glands. The stratum corneum of thick skin is about 300 µm, in contrast to an average of 14 µm in thin skin, where it is too thin to be visualized in detail by OCT[21]. In thin skin, epidermal thickness fluctuates between 70 µm to 120 µm, with the full thickness of the epidermis plus the dermis varying between 1000 µm to 2000 µm[19].

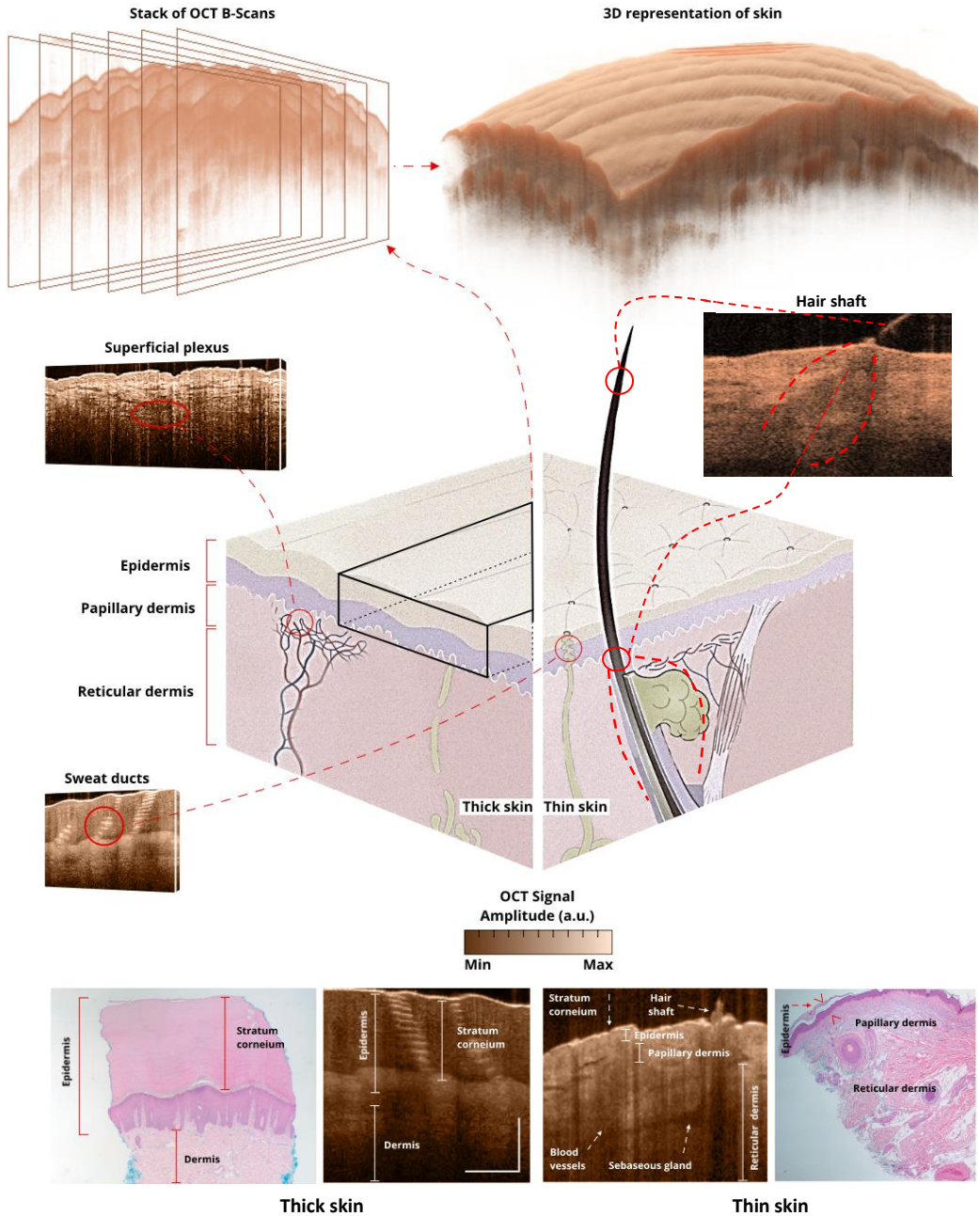

Figure 1. The illustration demonstrates the sequential images obtained by OCT (top left), and the 3D OCT representation of the skin (top right). The center illustration demonstrates several skin structures and their corresponding appearance on OCT. The bottom images demonstrate thick skin and thin skin, and annotated structures, their corresponding equivalent histology, and OCT images. The scale bar in OCT images is 400 µm.

Quantification of tissue cellular and architectural characteristics through extraction of optical and textural features of skin tissue can be utilized in the analysis of OCT images[22-24,25]. *Optical properties* describe cellular characteristics of skin tissue that can be extracted by solving the light-matter equation, using single or multiple scattering models[26], conjugated with some OCT image analysis algorithms. The single scattering model assumes that only the light undergoing single scattering (ballistic photons) preserves the coherence properties and contributes to the OCT signal. The multiple scattering model however, is based on the extended Huygens–Fresnel (EHF) principle where the shower curtain effect is taken into account[27]. Both models have been used for investigating optical properties of tissue. Among optical properties derived from OCT images, attenuation coefficient, defined as light intensity decay due to absorption and scattering, has been successfully used for clinical diagnosis and characterization of skin abnormalities and diagnosis[28,29]. Textural features are formed from the variation in back-scattered light returned from micro-compartments with different sizes and densities[23,]. *First-order texture* features are statistics calculated from the image histogram that measures the probability of a certain pixel occurring in the image, and do not consider pixel neighborhood relationships. To derive *second-order statistics*, the statistical texture features from the gray level co-occurrence matrix (GLCM)[30], the spatial relationship between two pixels, are considered. The GLCM tabulates the number of times different combinations of pixel pairs, of a specific gray level, occur in an image, in four different directions (0°, 45°, 90° and 135°). To derive *higher order statistics*, the statistical texture features from the gray level run length (GLRLM) matrix[31], the spatial relationship between more than two pixels, are considered. In a given direction, GLRLM measures the number of times there are runs of consecutive pixels with the same value.

Diagnosis of skin disease currently relies on the training, experience, visual perception and judgment of the clinician. Further diagnostic information is obtained from histologic interpretation of biopsies of tissue samples. Both visual and microscopic inspection of tissue rely on physicians analyzing visible patterns to guide the diagnosis. Issues arise when, for the same patient, dermatopathologists disagree on the clinical and histological diagnosis, due to variability in visual perception. Tools for automated pattern recognition and image analysis provide objective information to support clinical decision-making and may serve to reduce this variability. Previous studies have demonstrated that utilizing OCT techniques such as polarization-sensitive OCT in conjunction with advanced image analysis methods, healthy and neoplastic tissues, particularly basal cell carcinoma, can be differentiated[12,14,22,32]. However, in those studies typically qualitative and visual features[33] are used for structure identification. Other limitations of those studies include use of *in vitro* data and use of complex and expensive techniques such as polarization-sensitive OCT[22], using less efficient features, and/or using inefficient analysis methods. Other studies did not fully incorporate all available data acquisition and analysis techniques. In this manuscript, we propose a model based on analysis of optical and texture features to describe the gray-level patterns, pixel interrelationships, and the spectral properties of an image, in order to provide the objective analysis of tissue samples in a noninvasive manner. The aim of this study is to create comprehensive *in vivo* models of human skin diseases using numerical features extracted from OCT images and to use such models to assist in the diagnosis of common skin disorders. Our study is designed to be completed in two phases. In the *first phase,* optical and textural features extracted from OCT images of healthy skin at different body sites *in vivo* are analyzed and compared. In the *second phase,* the same features are extracted from OCT images of diseased skin and surrounding healthy tissue, these are used for computational modeling. The models will then be tested on diseased images to identify possible dermatological conditions.

**Dataset construction**

*Healthy skin in OCT images*: A stack of 170 images were taken for each of 10 body sites for each of the 10 healthy subjects, providing 17,000 images to develop a comprehensive analytical model of healthy tissue. The OCT B-scan images of nose, preauricular, volar forearm, neck, palm, back, thumb, dorsal forearm, sole, and calf were taken from male subjects aged between 25 and 52 years old, none of whom had any skin conditions. Among numerous images, collectively, the resulting 1000 images represented the data set for the first part of study. See Figure 2 for image examples. The images[32] were segmented into two-distinct layers using our semi-automatic DEJ detection algorithm that is based on graph theory. The algorithm was performed in an interactive framework by graphical representation of the attenuation coefficient map through a uniform-cost search method (see Figure Supplementary S3)[34]. The segmentation results were also verified by manual segmentation performed by two dermatologists.

*Diseased skin in OCT images*: The characteristics of diseased skin, hence the corresponding features in the OCT image, are altered compared to those of healthy skin. We studied epithelial skin tumors, i.e., basal cell carcinoma (BCC) and squamous cell carcinoma (SCC) for this study.

The diseased images in this study were taken from 11 subjects, aged between 25 to 52 years old, with histopathologically confirmed diagnosis of BCC or SCC. Each patient had one tumor; 5 with BCC and 6 with SCC. We collected 170, 2D images from each tumor at different transversal locations. We selected our sample images among those images. Although, we collected many images, some of them were excluded. One reason for exclusion was the inability to confirm with their histopathology match. Another reason was to have distinct SCC and BCC samples, in some cases SCC and BCC were very similar and cannot easily be distinguished. Our dermatologists (S.D & D.M) with histopathology expertise evaluated the OCT images and compared the results with biopsied tissue samples from that site, to identify the presence of BCC or SCC. For healthy images histology was not acquired. The images were manually (with the confirmation of histology images) annotated, generating 242 diseased skin images comprised of 119 BCC and 123 SCC images as our dataset. An additional 160 images were collected from locations at a minimum distance from the tumor that they could dermatologically be confirmed tumor-free, the other healthy iamges are collected from our healthy sample collected in *Healthy skin in OCT images* where sites were identical, totally 482 images. OCT images Based on histology results, our data set comprised nodular, superficial, and infiltrative subtypes of BCC and invasive SCC.

In Figure Supplementary S1, the OCT images and corresponding histology images for BCC are shown. In both the OCT and histology images of BCC, the central portion of the epidermis is ulcerated and covered with a crust (green arrow). SCC lesions develop from atypical cells with squamous cell characteristics proliferating in the dermis and underlying tissue. On the skin surface this appears as destruction of the epidermis, and local thickening of the tissue due to hyperkeratosis and disordered epidermal layering. Criteria used to determine SCC in OCT images were changes to tissue layer architecture and disruption of the basement membrane[35-37]. In Figure Supplementary S2, the OCT image and its corresponding histology image for an SCC sample are shown.

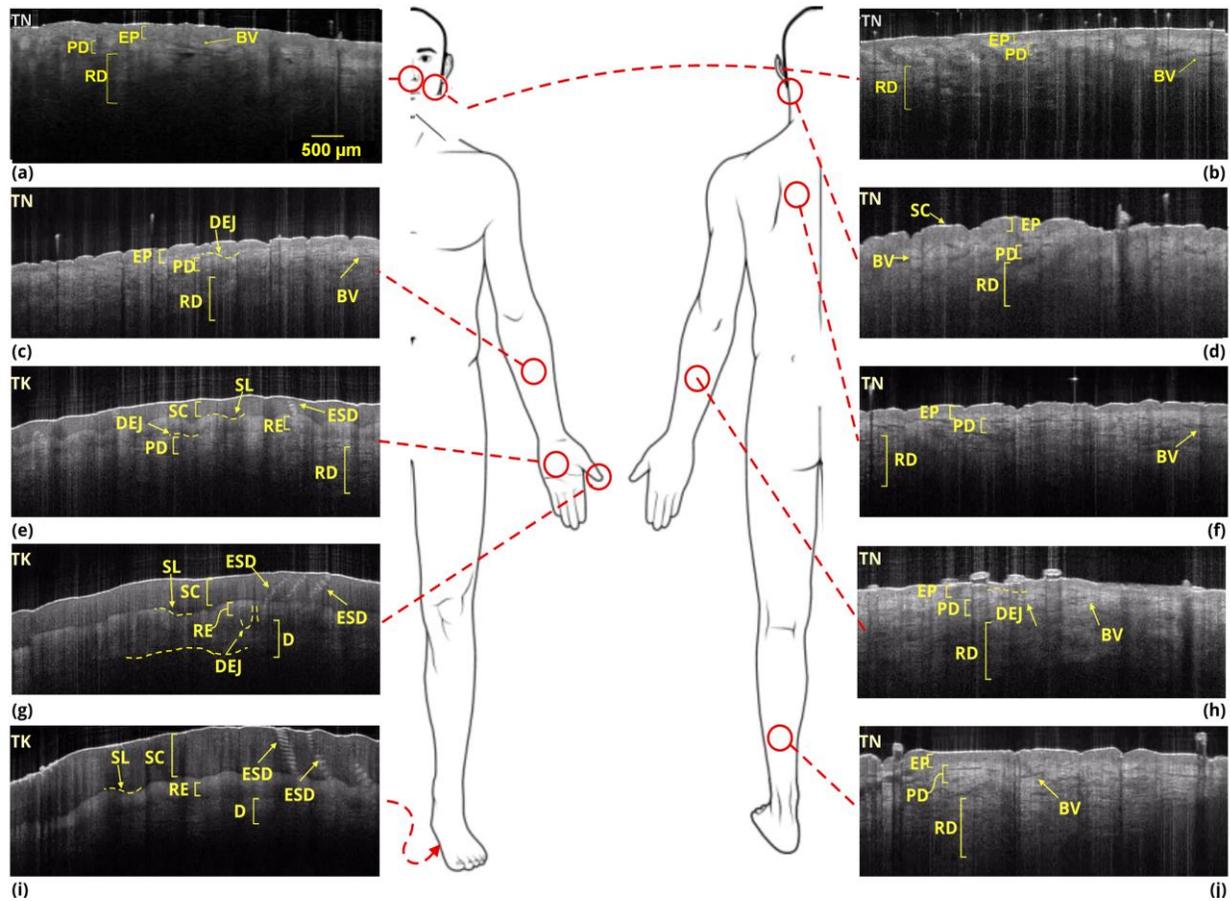

Figure 2. OCT images and structure annotation of different sites of body including, (a) nose, (b) preauricular, (c) volar forearm, (d) neck, (e) palm, (f) back, (g) thumb, (h) dorsal forearm, (i) sole, and (j) calf. SC: stratum corneum, SL: stratum lucidum, ESD: eccrine sweat ducts, RE: remainder of epidermis (stratum granulosum, stratum spinosum, stratum basale), RD: reticular dermis, DEJ: dermal-epidermal junction showing pronounced dermal papillae, PD: papillary dermis, D: dermis, EP: epidermis, BV: blood vessels, TK: thick skin, and TN: thin skin.

**Results**

Optical, first order statistical, and textural features, including sixty-three features, were extracted for both healthy and diseased image data sets.

*OCT Healthy skin* **:** These features are investigated and compared for both epidermis and dermis layers of healthy skin of patient's ten body sites. We observed that the value of these features varies but not sinificantly between skin of different sites due to the composition and arrangement of cells and organelles. We used ANOVA analysis (interval plots) to analyze the variation of the features for different sites of body and t-test to measure the inter-correlation between the features of the layers in both dermis and epidermis. A simple block diagram of the computational algorithm is depicted in Figure 5 (b) (Materials and Methods Section). *Optical features,* attenuation coefficient, is determined based on light intensity decay. Attenuation coefficient has been computed for different skin sites based on the single scattering calculation algorithm. The attenuation coefficient calculation algorithm is depicted in the block diagram in Figure 5 (a). In Figure Supplementary S4, attenuation coefficients of dermis and epidermis are shown for (a) nose, (b) preauricular, (c) volar forearm, (d) neck, (e) palm, (f) back, (g) thumb, (h) dorsal forearm, (i) sole, and (j) calf of ten healthy individuals. We observed that the palm and thumb are closely correlated in terms of attenuation coefficient. The attenuation coefficient is significantly different between the group of sole, palm and thumb compared to the other sites of body ($p < 0.05$) in both dermis and epidermis. Variation is also observed between preauricular and other sites for both dermis and epidermis. For the dermal layer, differences were detected between the sole and nose as well as between the sole and volar forearm. Figure Supplementary S4 also shows the map of *p*-values for epidermis and dermis of different body sites. *First-order statistical features (FOS)* extracted from the OCT images, were mean, standard deviation, variance, skewness, kurtosis, median and entropy. We observed insingificant differences for all FOS features extracted from epidermis and dermis layers in all skin sites. Figure Supplementary S5 and S6 show standard deviation and entropy as well as maps of their difference between each pair of skin sites. It was noted that the epidermal layer of nose and preauricular skins have similar texture. The results show that FOS features have the same trend as the optical property but also provide a more significant differentiation for both epidermis and dermis. *Textural features*, GLCM textures including contrast, energy, correlation, homogeneity and entropy in four directions (0°, 45°, 90° and 135°) were computed for all skin sites. The GLCM values of contrast and entropy at 45 degrees for epidermis and dermis as well as their maps of *p*-value are illustrated in Figure Supplementary S7 and S8. Our findings show that there is a significant difference, $p < 0.05$, between the textural features of different body sites with different epidermal thickness, i.e thick and thin skin types in most cases. Although not shown, all optical, statistical, textural features, were computed for epidermis and dermis. We also observed a neligible difference between the *GLRLM textural features* including short run emphasis (SRE), long run emphasis (LRE), gray-level nonuniformity (GLN), run percentage (RP), run length nonuniformity (RLN), low gray-level run emphasis (LGRE), high gray-level run emphasis (HGRE) that are also computed for all skin sites of different body sites. Oveal, the Textures are not higly differentiable for the skin from different body sites. This was not unexpected because they are all healthy skin samples. The analysis shows FOS features for different sites of body are almost in the same range for all body sites.

*Classification:* Together, the resulting 482 images represented the training and testing set of the classification problem described in the following section. The sequential images of the diseased skin of 11 subjects were manually (with the confirmation of histology images) annotated, generating 242 diseased skin images comprised of non-melanoma skin cancers including 119 basal cell carcinoma (BCC) and 123 squamous cell carcinoma (SCC) images as our dataset. A region of interest (ROI) for a given B-scan pixels in each image was chosen such that the tumorous region is selected. In addition to the corresponding healthy skin images of identical sites collected in first phase, other ROIs were also chosen from the surrounding healthy tissue of tumor. The optical, statistical, and textural features were extracted from the images. Different combinations of features were evaluated in classification algorithms. A principle component analysis (PCA)-based feature selection[38] method was used resulting in the selection of six features: one optical, three statistical and two textural features. These features are attenuation coefficient, entropy from FOS, entropy at 0 degrees, correlation at 0 degrees, correlation at 135 degrees and homogeneity at 135 degrees from GLCM. The results pertaining to differentiating BCC from healthy skin as well as SCC from healthy skin are shown in Figure 3 and Figure 4, respectively. Figure 3 (a) and Figure 4 (a) depict the correlation between each two pair of features in the 63-feature pool. Brighter colors indicate larger correlation value. A noticeable correlation is observed among the features from the same category. Figure 3(b) and Figure 4(b) show that the remaining features (after feature selection) have less correlation with each other. Several machine learning classifiers were tested for solving the classification problem. Among all, we showed the results of SVM[39] with two different kernels, i.e., linear and 2$^{nd}$ polynomial (quadratic), logistic regression[40], K-nearest neighbor classifiers[41], linear discriminant analysis (LDA)[42] and artificial neural network (ANN)[43] that had higher accuracy. Linear SVM (LSVM) yielded the optimum result with an accuracy rate of 80.9, specificity of 80.5 and sensitivity of 81.9 for BCC classification. With quadratic SVM (QSVM) an accuracy rate of 87.2%, specificity of 87.3 and sensitivity of 87 were obtained for SCC classifications. QSVM also provided satisfying results, 80.5% accuracy for BCC classification. These results were obtained when the classifier was used with the six selected features.

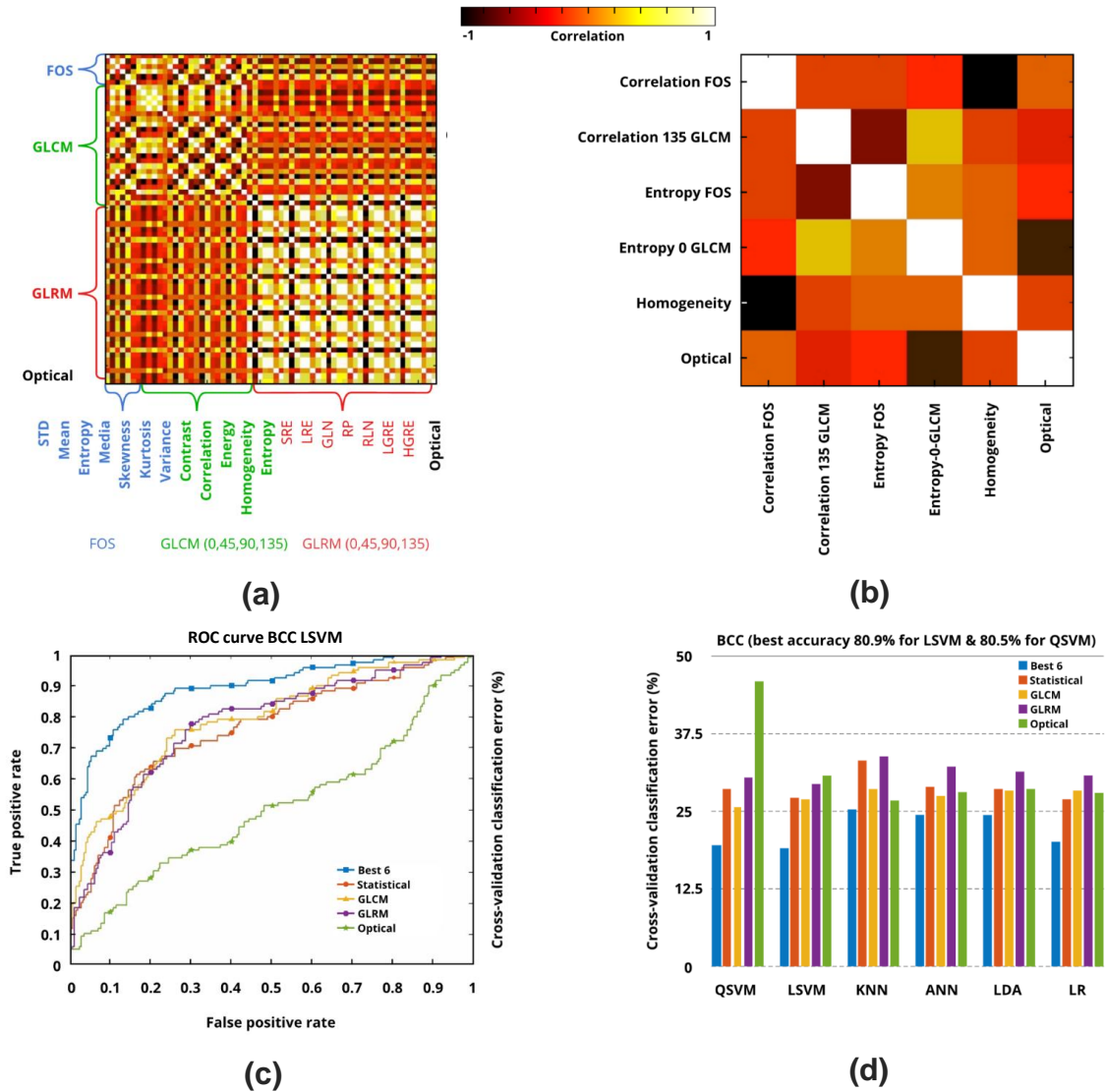

Figure 3. Classification results. (a) Correlation map of 63 features used for differentiating healthy versus BCC samples, (b) correlation map of the six selected features, (c) ROC curve for different subsets of features with LSVM classifier, (d) cross-validation classification error of different classifiers when different subsets of features were used. SRE: short run emphasis, LRE: long run emphasis, GLN: gray-level nonuniformity, RP: run percentage, RLN: run length nonuniformity, LGRE: low gray-level run emphasis, HGRE: high gray-level run emphasis.

The receiver operating characteristic (ROC) curve for several subsets of features with LSVM and QSVM as classifier are shown in Figure 3(c) and Figure 4(c), respectively. Figure 3(d) and Figure 4(d) show the cross-validation classification error percentage of the six classifiers when different subsets of features are used.

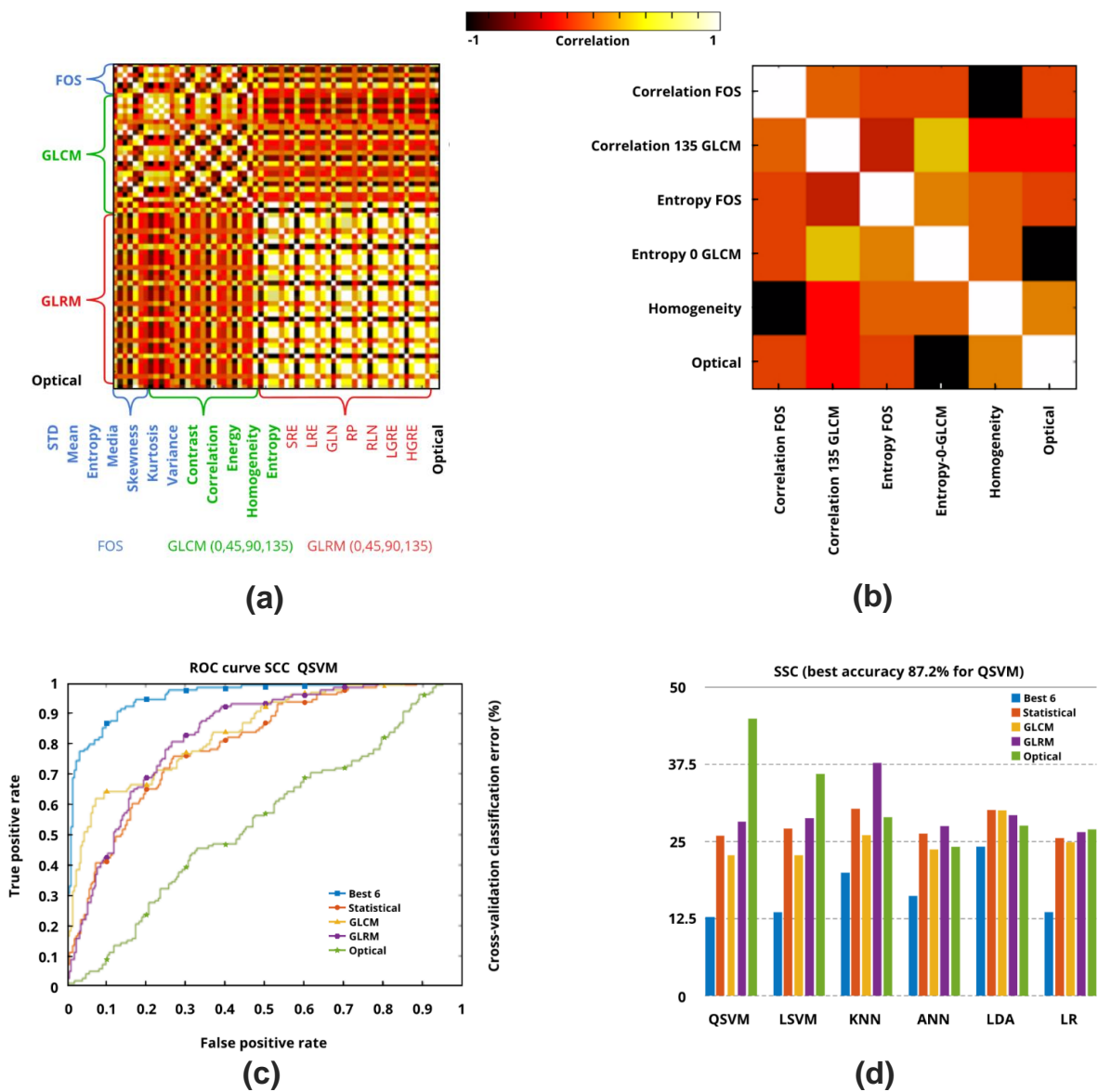

Figure 4. Classification results. (a) Correlation map of 63 features used for differentiating healthy versus SCC samples, (b) correlation map of the six selected features, (c) ROC curve for different subsets of features with QSVM classifier, (d) cross-validation classification error of different classifiers when different subsets of features were used. SRE: short run emphasis, LRE: long run emphasis, GLN: gray-level nonuniformity,

RP: run percentage, RLN: run length nonuniformity, LGRE: low gray-level run emphasis, HGRE: high gray-level run emphasis.

**OCT versus high resolution ultrasound**

High-frequency ultrasound is mainly used to estimate tumor thickness in melanoma, to plan one-step excisions with appropriate margins, and help to determine the necessity of sentinel lymph node biopsy[4]. Its penetration depth lies around 8 mm at 20 MHz. We imaged the skin of the same body sites with several OCT and ultrasound imaging systems in order to compare their resolutions and penetration depths. The modalities used were a swept source OCT (SS-OCT), clinical ultrasound (9 MHz), high frequency (HF) ultrasound (48 MHz), ultra-high frequency (UHF) ultrasound (70 MHz) and high definition (HD) OCT. These images are shown in Figures Supplementary S9 to S12, and their histology images given in Figure Supplementary S13. The speckle size in OCT and ultrasound images of a fabricated tissue-mimicking phantom, composed of $TiO_2$ and polyurethane, were listed in Table 1 for comparison. Average speckle size is estimated by using the full width at half maximum (FWHM) of the auto-covariance function of the speckle pattern[44]. In Table 1, we also compared the resolution, field of view and penetration depth of these imaging modalities. Comparing the results, OCT surpasses other modalities in terms of speckle size. SS-OCT is the most favorable one due to its moderate penetration depth, resolution, field of view, and speckle size.

Table.1 Specifications of the imaging modalities used in this paper.

| Imaging modality | Model | Axial resolution (μm) | Lateral resolution (μm) | Image size (width (mm) × height (mm)) | Averaged speckle size for a phantom (μm) |
|---|---|---|---|---|---|
| SS-OCT | Vivosight | 7.5 | 10 | 6 × 2 | 22 |
| HD OCT | Agfa Skintell | 3 | 3 | 1.8 × 1.5 | 15 |
| Clinical US | Sonoscape S9 BW: 15-5 MHz | 100 | 200 | --- | 400 |
| US_UHL | 48UHF Vevo MD BW: 20-46 MHz | 50 | 110 | 15.4 × 23.5 | 225 |
| US_UHR | 70UHF Vevo MD BW: 29-71 MHz | 30 | 65 | 9.7 × 10.0 | 114 |

**Discussion**

OCT is an effective imaging modality capable of aiding in the diagnosis of skin conditions including inflammatory diseases and non-melanoma skin cancer. The diagnosis of skin disease is based primarily on the visual assessment of the dermatologist and recognizing patterns of morphology. Noninvasive skin imaging techniques, including OCT, can provide further information to the clinician. Currently clinicians rely on their visual pattern recognition skills and expertise as a physician viewing the images. Tools for automated pattern recognition and image analysis can provide objective information to support clinical decision-making. This study presents the incorporation of clinical and detailed quantitative textural assessment of OCT images to first generate a comprehensive morphological and computational atlas of normal human skin *in vivo*. The reference system of *in vivo* healthy skin OCT images can then be used to assess a wide variety of skin disorders with the aim of improving diagnosis. We generated a small-scale OCT atlas of human skin from sites shown in Figure 2 (nose, preauricular, volar forearm, neck, palm, back, thumb, dorsal forearm, sole, and calf), which covers variations of skin tissues throughout the body. We imaged healthy skin from a variety of body sites from different individuals. The images were then segmented using our dermal-epidermal junction (DEJ) detection algorithm. The algorithm is based on graph-theory representation of the attenuation coefficient map through a uniform-cost search method. Features including attenuation coefficient, statistical, and textural features were extracted from ten evenly distributed ROIs in both epidermis and dermis of different body sites. The average values and their corresponding 95% confidence interval (CI) across different skin sites were calculated. The derived features were different for the dermis and epidermis in healthy skin of different sites. These features were then extracted from OCT images of diseased and healthy skin and used for classification.

The epidermis and dermis vary in different anatomic areas. Optical properties and hence the corresponding numerical features in OCT images vary based on sizes, shapes, concentration and orientations of tissue microstructure; cell membranes and blood vessel walls act as scatterers / reflectors and refractors. In texture analysis, the attribute 'contrast' of the GLCM is a measure of texture analysis, showing the difference between the highest and lowest intensity values of a set of pixels. This parameter was significantly different between the values calculated from palm/sole and nose. The attribute 'energy' of the GLCM matrix is a measure of uniformity of pixel pair recurrences and identifies disorders in texture. High-energy values occur when gray level distribution has a constant or periodic form. Significant variations of energy were measured in sole samples as compared to all other sites for both the epidermis and dermis. In the case of the attribute 'entropy' of the GLCM, we have an identifier of disorder or complexity of an image that is large when the image is not texturally uniform. Sole, palm and thumb showed a significant difference in entropy when compared to that of other sites in both dermis and epidermis. The attribute 'inverse difference moment' or 'homogeneity' of the GLCM, in spite of having dissimilarity, did not offer a significant distinction among different sites. With the numerical features extracted from OCT images, we successfully trained a classifier to differentiate between healthy and abnormalities of dermal microstructure. Among the classifiers we examined, SVM offered the best accuracy to differentiate between normal and abnormal tissue samples. In this workflow, we used an efficient, limited number of features and a modified PCA algorithm for feature selection. Thus, our algorithm might be limited as result of PCA limitations[34]. Although this selection of features covers an adequate variety in the projected space, their values may not linearly (or quadratically) discriminate between two classes. Therefore, future directions for research include, a larger data set, exploring other efficient features, and investigation of more efficient feature selection and classification algorithms. Based on our data analysis in terms of recall and perception, it is observed some examples where the propsed classifier has failed and BCC or SCC skin tissue were assessed as non-cancerous by proposed workflow. The reason for this classifier misinterpretation may be due to similarity of the cancerous tissue to surrounding texture; see example Figure Supplementary S14 (a) and (b). This objectively determined information could assist clinicians to diagnose, develop treatment plans, and determine individual prognoses more accurately. Currently high frequency ultrasound is the most commonly utilized cutaneous imaging modality. Theoretically, some of high frequency ultrasound systems have a resolution close to that of OCT or even better. We however observed more distinct structures in OCT images. Comparing the results given in Table 1, OCT surpasses other modalities in terms of speckle size. SS-OCT is the most favorable due to its moderate penetration depth, resolution and field of view.

It is worth mentioning that we are using a methodology where we do not need to decorrelate the effect of confocal gate and sensitivity drop-off since the peak of the confocal and coherence gates move simultaneously. Similarly, with some modifications, our Vivosight OCT can be configured to approximate the system as a discrete dynamic focus OCT and, with a good approximation, ignore these parameters[16].
Therefore, compensation for confocal parameter of the lens and for the fall in laser coherence with distance from zero path length difference was not performed. However, both of these effects can be safely neglected over the imaging depth of interest < 1.0 mm due to the multi-beam optics of the OCT device we used[45,46].

In summary, we have extracted optical, textural, and statistical properties from OCT healthy skin images to create a computational atlas of the normal skin at different anatomic sites. We observed that skin cellular architecture varies across the body, and so do the textural and morphological characteristics in the OCT images. There is, therefore, a critical need to systematically analyze OCT images of different sites and identify their significant qualitative and quantitative differences. We demonstrated that the computational models can assist in diagnosis of abnormalities of dermal microstructure, i.e., BCC vs. healthy, or SCC vs. healthy, and hence aid in the determination of treatment. The proposed workflow can be generalized for detection of other tissue abnormalities. The result of this study can be extended as an interactive machine learning kernel interface addable to OCT devices.

**Materials and Methods**

**Participants and dataset**

OCT images were taken from ten subjects, aged between 25 to 52 years old none of whom had any skin condition. For each subject, specific regions of nose, preauricular, volar forearm, neck, palm, back, thumb, dorsal forearm, sole, and calf were imaged and analyzed. A specialized holder is used for the OCT probe to make sure that we consistently imaged the same area of skin on each subject. For the disease classification part, images in this study were taken from 11 subjects, aged between 25 to 52 years old, with histopathologically confirmed diagnosis of BCC or SCC. Each patient had one tumor; 5 with BCC and 6 with SCC. We collected 170, 2D images from each tumor. We selected our sample images from among those images. Although, we collected many images, some samples were excluded. One reason for exclusion was the inability to confirm with their histopathology match. Another reason was to have distinct SCC and BCC samples, in some cases SCC and BCC are very similar and cannot easily be distinguished. Our dermatologists (S.D & D.M) with histopathology expertise evaluated the OCT images and compared the results with biopsied tissue samples from that site, to identify the presence of BCC or SCC. For healthy images histology was not acquired. An additional 240 images were collected from locations at a minimum distance from the tumor that they could dermatologically be confirmed tumor-free as well as healthy imae set Based on histology results, our dataset comprised nodular, superficial, and infiltrative subtypes of BCC and invasive SCC. We added a new section, Dataset, to the revised manuscript to provide a more detailed explanation about the images we used and how they were generated and selected. All imaging procedures were carried out according to the guidelines of the US National Institutes of Health, and Institutional Review Board (IRB) approval board of the Wayne State University and informed consent was obtained from all subjects before enrollment in the study. Images for the skin conditions were collected in the Wayne State University Physician Group Dermatology Clinic, Dearborn, MI.

**OCT system**

Figure Supplementary S15 shows the schematic illustration of the multi-beam, swept-source OCT (SS-OCT) system (Vivosight, Michelson Diagnostic ™ Inc., Kent, United Kingdom), used in this study. The lateral and axial resolutions are 7.5 µm and 10 µm, respectively. The scan area of the OCT system is 6 mm width by 6 mm length by 2 mm depth. A tunable broadband laser source (Santec HSL-2000-11-MDL), with the central wavelength of 1305 +/- 15 nm successively sweeps through the optical spectrum and leads the light to four separate interferometers and forms four consecutive confocal gates. The 10 KHz sweep is the frequency of generating one reflectivity profile (A-Scan). A B-Scan is then generated by combining several adjacent A-Scans for different transversal positions of the incident beam. B-scan frame rate is 20 frame/s[47].

**Data Analysis.**

Healthy OCT images of skin are first segmented into two distinct layers using our semi-automatic DEJ detection algorithm[34]. The algorithm works based on converting a border segmentation problem to a shortest-path problem using graph theory. It is performed in an interactive framework by graphical representation of an attenuation coefficient map through a uniform-cost search method. To smooth borders, a fuzzy algorithm is introduced enabling a closer match to manual segmentation. The details of this method have been reported previously[34]. The diseased parts of the OCT image are manually selected based on the histopathology images. The ROI was selected such that the tumorous region is within it. ROIs from the surrounding healthy skin were also chosen. The images then go through the procedure depicted in Figure 5 (b), where the optical, statistical and textural features are extracted from the images. To suppress the speckle noise, a BM3D filter was used[48]. The despeckled images were used for better visualization as well as segmentation.

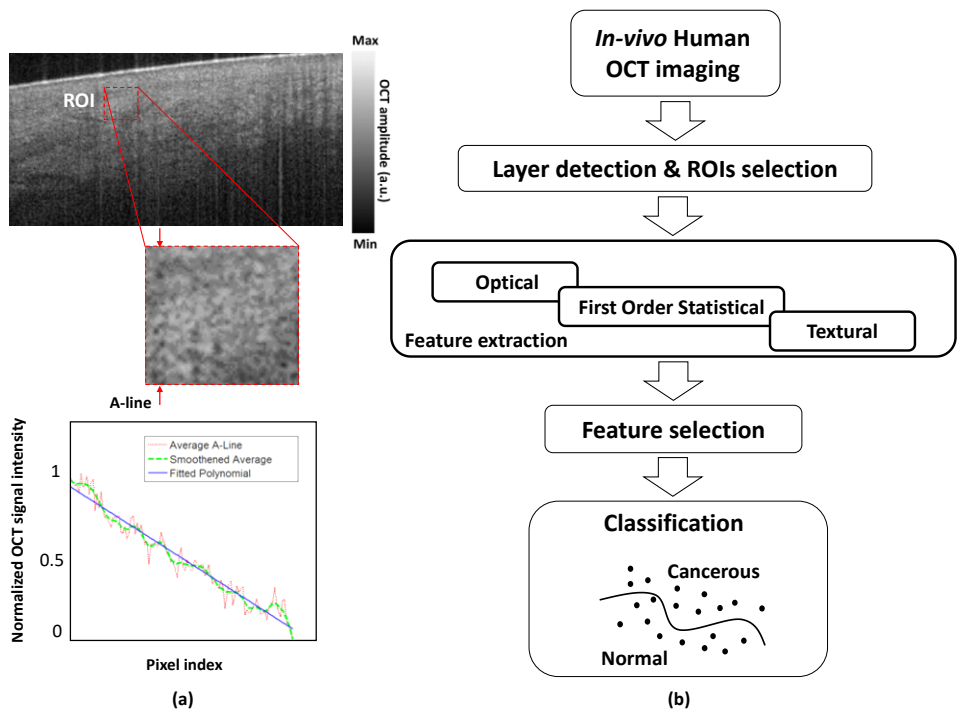

Figure 5. (a) Attenuation coefficient measurement procedure, (b) block diagram of the proposed computational method.

*Optical feature:* We calculated the attenuation coefficient as the optical property of the tissue. The A-scans in each ROI were averaged. The Levenberg-Marquardt algorithm was used for curve-fitting. The attenuation coefficient of the ROI in the sample was then the slope of the fitted curve on the averaged A-scan (see Figure 5 (a)). *First-order statistical features:* Mean, variance, standard deviation, skewness, median, entropy and kurtosis were calculated for each ROI. First-order measures are statistics calculated from the original image values, and do not consider pixel neighborhood relationships. They are computed based on the intensity value concentrations on all or part of the histogram. *Second-order statistical features:* We used statistical texture features from the gray level co-occurrence matrix (GLCM) [30] to represent second-order statistics. These features demonstrate the spatial relationship between pairs of pixels. The GLCM tabulates the number of times different combinations of pixel pairs of a specific gray level occur in an image in different directions. Homogeneity, contrast, energy, entropy and correlation in four directions, 0°, 45°, 90° and 135°, are calculated as second-order statistics. *Higher order statistical features:* We used statistical texture features from the gray level run length (GLRLM) matrix to represent higher order statistics. These features demonstrate spatial relationship between more than two pixels. In a given direction, GLRLM measures the number of times there are runs of consecutive pixels with the same value including short run emphasis (SRE), long run emphasis (LRE), gray-level nonuniformity (GLN), run percentage (RP), run length nonuniformity (RLN), low gray-level run emphasis (LGRE), high gray-level run emphasis (HGRE) [31]. We constructed a feature vector comprised of FOS textures, GLCM textures, and GLRLM features in four angular directions, 0°, 45°, 90° and 135°. The mean of the obtained features for dermis and epidermis and their corresponding 95% confidence intervals (CI) across different skin sites were estimated. ANOVA analysis (interval plots) was used to analyze the variation of these features for different sites of body in both dermis and epidermis. The differences in image features between sites were compared using t-test. We used Minitab Statistical Software (version 17.0, Minitab Inc., Pennsylvania, USA) for ANOVA analysis.

**Classifiers.**

Prior to classification, features were normalized, then a feature selection algorithm was performed to obtain the most discriminative features. We used principal component analysis (PCA) as our feature selection method. PCA finds a linear map from the data in a high dimensional space to a desired low dimensional space trying to preserve the data variance[41]. To perform PCA, we obtained the principal components and then kept the features which provided the greatest contribution to the first sixth principal components. After feature selection was performed, the images we had collected to fill the learning database were classified using machine learning classifiers. We tested SVM (with two different kernels of linear, LSVM, and 2$^{nd}$ degree polynomial (QSVM)), logistic regression (LR), k-nearest neighbor (KNN), linear discriminant analysis (LDA) and artificial neural networks (ANN). It has been shown previously that although SVM is designed to solve linear classification tasks, by using some kernel tricks, it can be used for nonlinear classification tasks and is very well suited for binary (two class) problems[39]. In LR classification, the probability that a binary target is true is modeled as a logistic function of a linear combination of features[40]. For (KNN) the rule classifies each unlabeled sample by the majority label among its K-nearest neighbors in the training set[41]. LDA, searches for a linear combination of variables that best separates binary targets[42]. An ANN classifier consists of many neurons, i.e., highly interconnected processing components, that work constructively and coherently to solve specific problems[43]. Classifiers were validated using $10 \times 10$-fold cross-validation method. In 10-fold cross-validation, the data is randomly split into 10 equal folds. The classification procedure is implemented in an iterative manner. For each run nine folds are used for training and one-fold is used for testing. The process is repeated ten times and the final accuracy is the average of all the fold accuracies.

**Implementation**. The approaches described in this study have been implemented in Matlab ® 2016 except the segmentation algorithm which is developed in Delphi. The experiments were carried out on a standard computer workstation (3.10 GHz Intel Core i7, 32 GB RAM). In addition to custom routines and semi-automatic ROIs selection developed by the authors using Matlab's built-in functions, publicly available source code for BM3D has been utilized[44].

**Acknowledgments**

This project has been funded by Michelson Diagnostics and by Wayne State University Startup fund. We acknowledge Dr. Gerald Hish from Wayne State University Laboratory Animal Resources for his help in using their clinical ultrasound imaging system. We also give a special thanks to VisualSonics Inc., Toronto, CA, for use of their high resolution ultrasound imaging system, Vevo MD. We also thank reviewer's for their constructive comments.

**Authors Contributions**

M.A and S.A designed the research. S.A, S.D, D.M and M.A. conducted the experiments. S.A, M.H, S.N, and M.A and S.C analyzed the data. S.A, and M.A, wrote the manuscript and all authors participated in paper revisions.

**Competing financial interests:**

The authors declare no competing financial interests.

**Figure captions**

**Figure 1.** The illustration demonstrates the sequential images obtained by OCT (top left), and the 3D OCT representation of the skin (top right). The center illustration demonstrates several skin structures and their corresponding appearance on OCT. The bottom images demonstrate thick skin and thin skin, and annotated

structures, their corresponding equivalent histology, and OCT images. The scale bar in OCT images is 400 µm.

**Figure 2**. OCT images and structure annotation of different sites of body including, (a) nose, (b) preauricular, (c) volar forearm, (d) neck, (e) palm, (f) back, (g) thumb, (h) dorsal forearm, (i) sole, and (j) calf. SC: stratum corneum, SL: stratum lucidum, ESD: eccrine sweat ducts, RE: remainder of epidermis (stratum granulosum, stratum spinosum, stratum basale), RD: reticular dermis, DEJ: dermal-epidermal junction showing pronounced dermal papillae, PD: papillary dermis, D: dermis, EP: epidermis, SG: sebaceous glands, BV: blood vessels, TK: thick skin, and TN: thin skin.

**Figure 3.** Classification results. (a) Correlation map of 63 features used for differentiating healthy versus BCC samples, (b) correlation map of the six selected features, (c) ROC curve for different subsets of features with LSVM classifier, (d) cross-validation classification error of different classifiers when different subsets of features were used. SRE: short run emphasis, LRE: long run emphasis, GLN: gray-level nonuniformity, RP: run percentage, RLN: run length nonuniformity, LGRE: low gray-level run emphasis, HGRE: high gray-level run emphasis

**Figure 4.** Classification results. (a) Correlation map of 63 features used for differentiating healthy versus SCC samples, (b) correlation map of the six selected features, (c) ROC curve for different subsets of features with QSVM classifier, (d) cross-validation classification error of different classifiers when different subsets of features were used. SRE: short run emphasis, LRE: long run emphasis, GLN: gray-level nonuniformity, RP: run percentage, RLN: run length nonuniformity, LGRE: low gray-level run emphasis, HGRE: high gray-level run emphasis.

**Figure 5.** (a) Attenuation coefficient measurement procedure, (b) block diagram of the proposed computational method.

# Supplementary materials

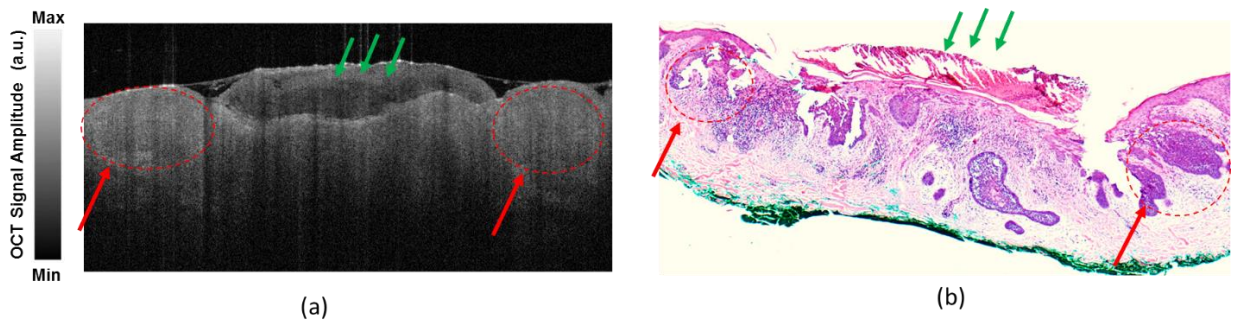

Figure Supplementary S1. (a) OCT B-scan, and (b) its corresponding histological image of a BCC from a 62 year old female. In both OCT and histology images of BCC, the central portion of the epidermis is ulcerated and covered with a crust (green arrow). On either side of the ulceration, there are tumor nodules (red arrows). On the histology images, there are artefactual fractures within the tumor masses that occurred during tissue processing so they are not present in the OCT images.

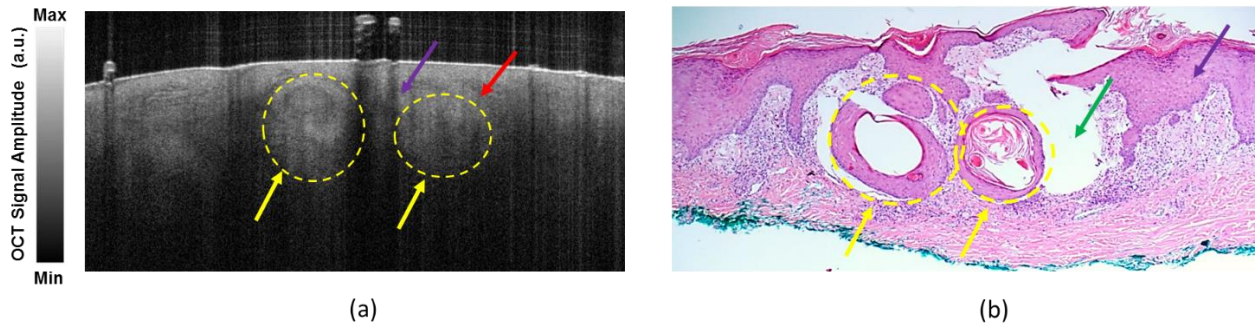

Figure Supplementary S2. (a) OCT B-scan, and (b) its corresponding histological image of an SCC from a 51 year old female. Keratinous pearls within SCC are shown with yellow arrows in both OCT and corresponding histology image. There is a proliferation of keratinocytes in the epidermis pushing into the dermis (red arrow). Keratinocytes in the epidermis show atypia (purple arrow) with large nuclei. The green arrow in the histology image, labels a tear in the tissue that we do not see on the OCT image since it is an artefact of tissue processing. On OCT image, the infiltration of tumor cells into the dermis leads to a loss of the dark line representing the dermo-epidermal junction.

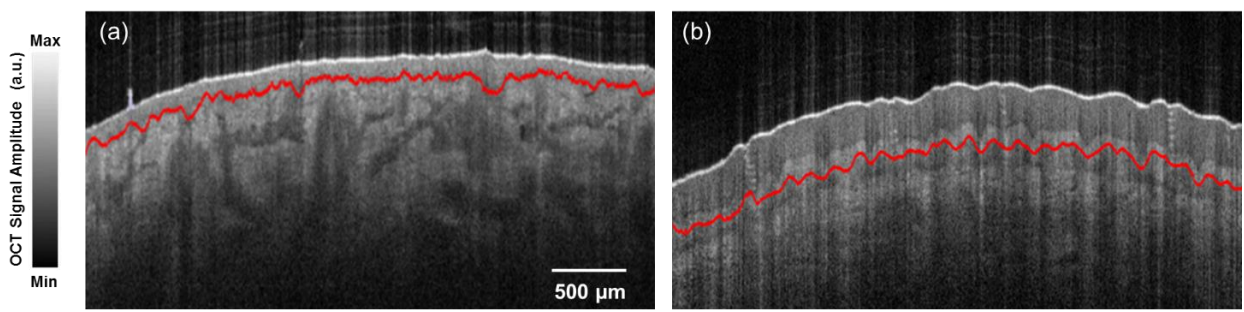

Figure Supplementary S3. Examples of the skin layer detection algorithm applied to OCT images. Application of our DEJ detection method on OCT images of (a) preauricular (thin skin) and (b) palm (thick skin). The red color is the dermal-epidermal junction obtained from our DEJ detection algorithm.

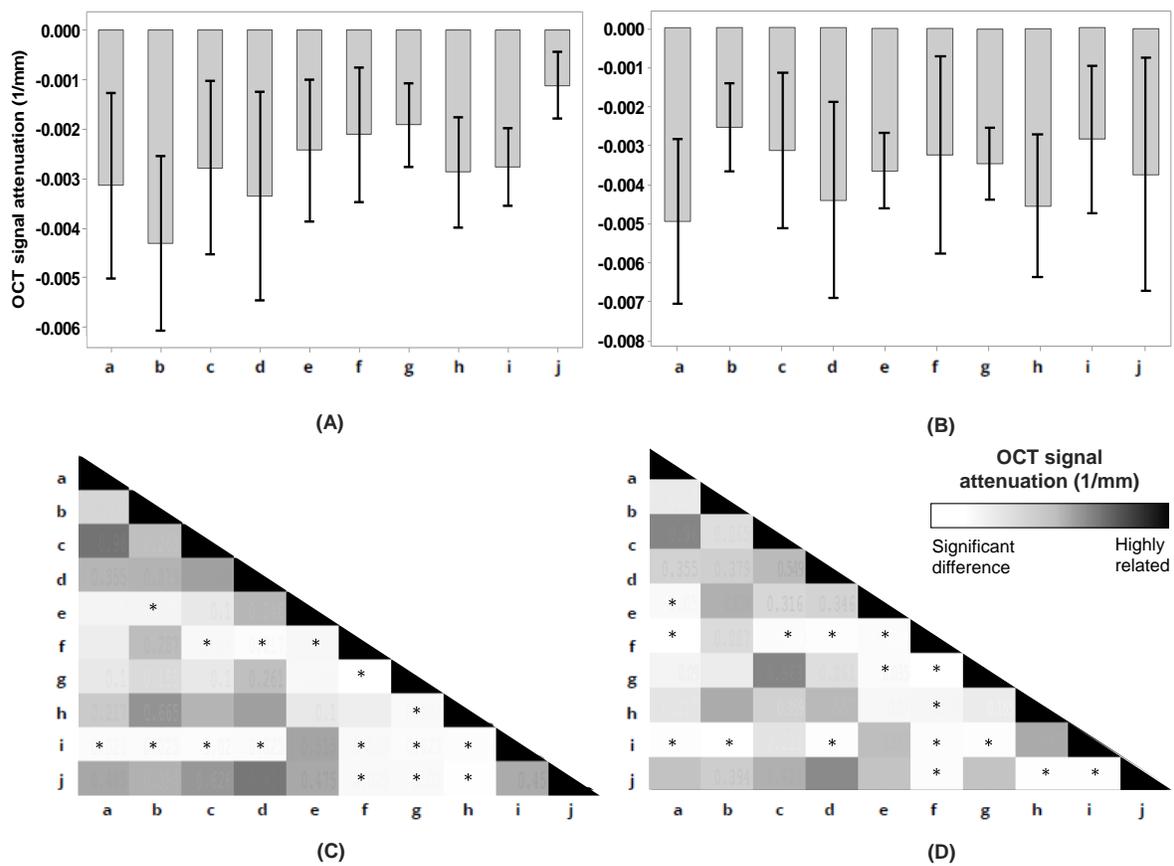

Figure Supplementary S4. Signal attenuation calculated for (A) epidermis, and (B) dermis of ten body sites; Calculated gray-scale coded map of *p*-values between each pairs of body site (*p*-value < 0.05 considered as an acceptable significant difference indicated with *) for (C) epidermis, and (D) dermis. The letters from a-j demonstrate the following; (a) tip of nose, (b) preauricular, (c) volar forearm, (d) neck, (e) palm, (f) back, (g) thumb, (h) dorsal forearm, (i) sole, (j) calf.

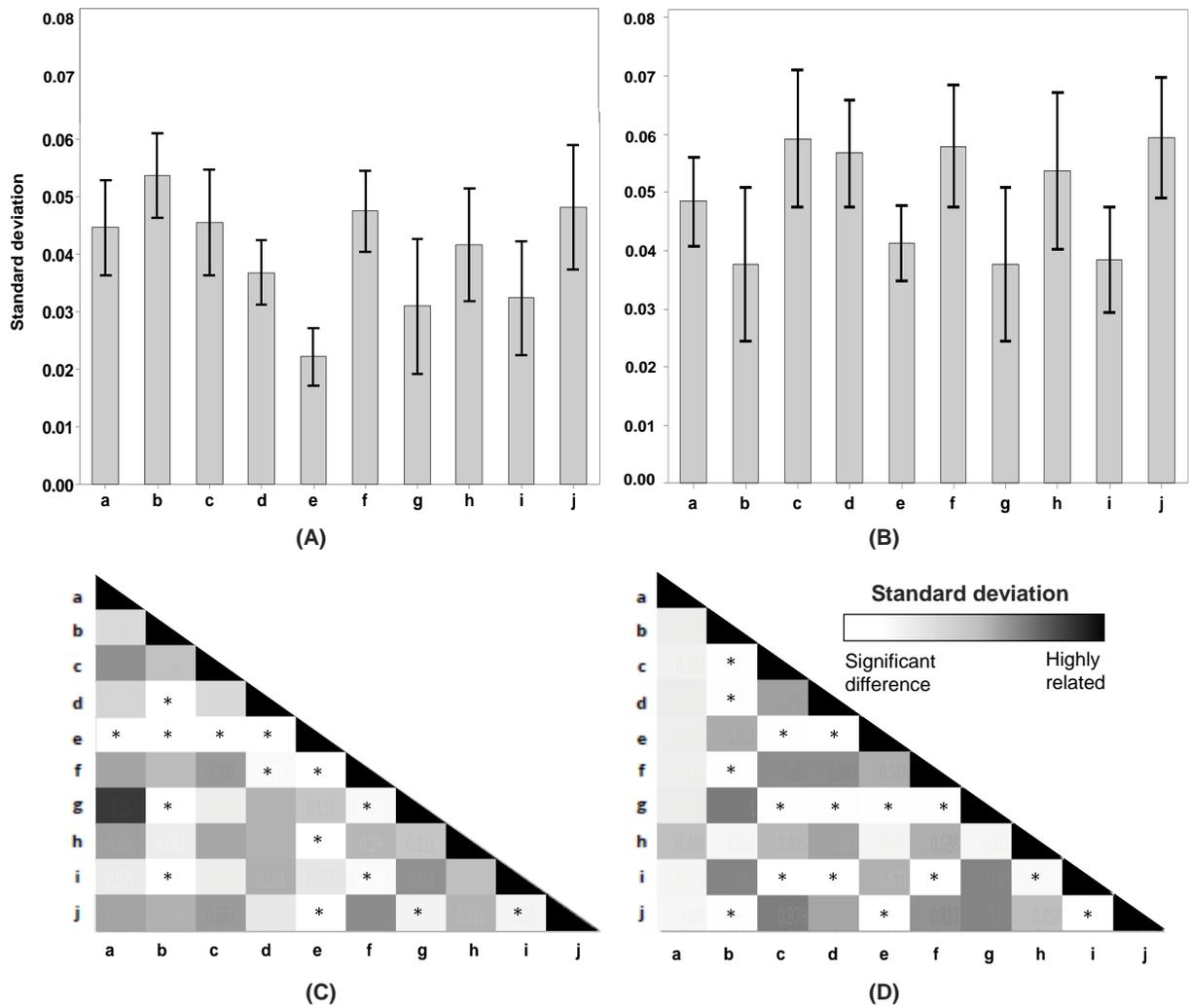

Figure Supplementary S5. First order statistical analysis, standard deviation, results and comparison for (A) epidermis, and (B) dermis of ten body sites; Calculated gray-scale coded map of *p*-values between each pairs of body site (*p*-value < 0.05 considered as an acceptable significant difference indicated with *) for (C) epidermis, and (D) dermis. The letters from a-j demonstrate the following; (a) tip of nose, (b) preauricular, (c) volar forearm, (d) neck, (e) palm, (f) back, (g) thumb, (h) dorsal forearm, (i) sole, (j) calf.

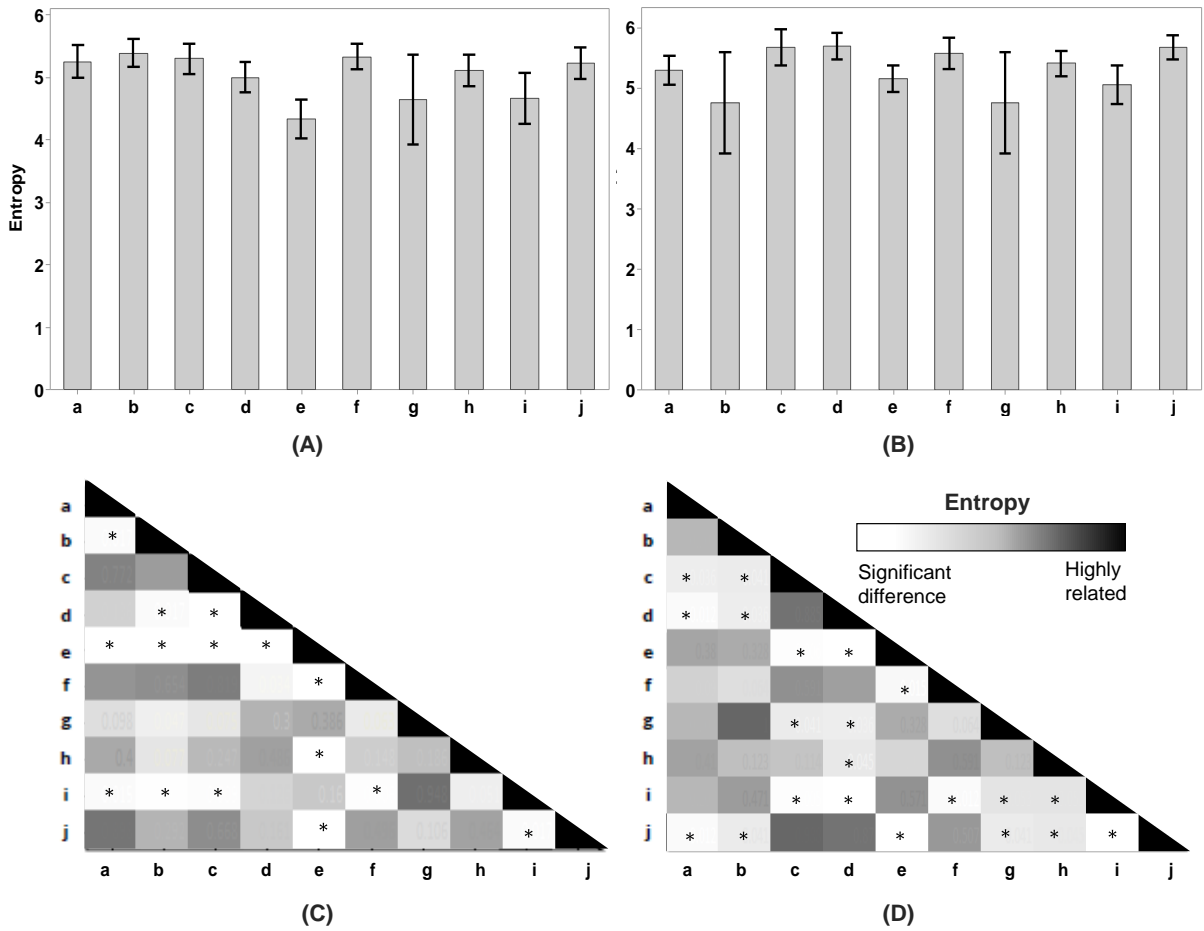

Figure Supplementary S6. First order statistical analysis, entropy, results and comparison for (A) epidermis, and (B) dermis of ten body sites; Calculated gray-scale coded map of $p$-values between each pairs of body site ($p$-value $< 0.05$ considered as an acceptable significant difference indicated with *) for (C) epidermis, and (D) dermis. The letters from a-j demonstrate the following; (a) tip of nose, (b) preauricular, (c) volar forearm, (d) neck, (e) palm, (f) back, (g) thumb, (h) dorsal forearm, (i) sole, (j) calf.

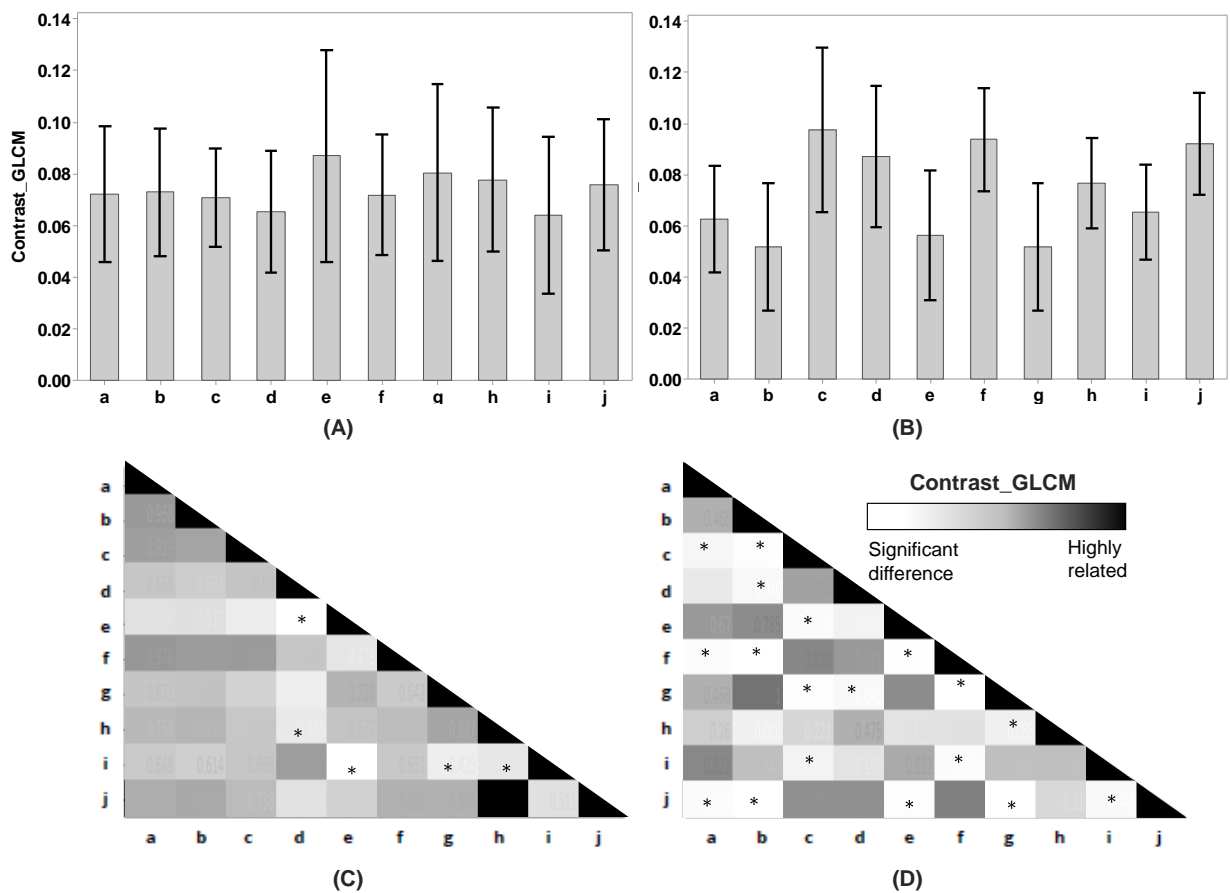

Figure Supplementary S7. GLCM texture analysis, contrast, results and comparison for (A) epidermis, and (B) dermis of ten body sites; Calculated gray-scale coded map of $p$-values between each pairs of body site ($p$-value $< 0.05$ considered as an acceptable significant difference indicated with *) for (C) epidermis, and (D) dermis. The letters from a-j demonstrate the following; (a) tip of nose, (b) preauricular, (c) volar forearm, (d) neck, (e) palm, (f) back, (g) thumb, (h) dorsal forearm, (i) sole, (j) calf.

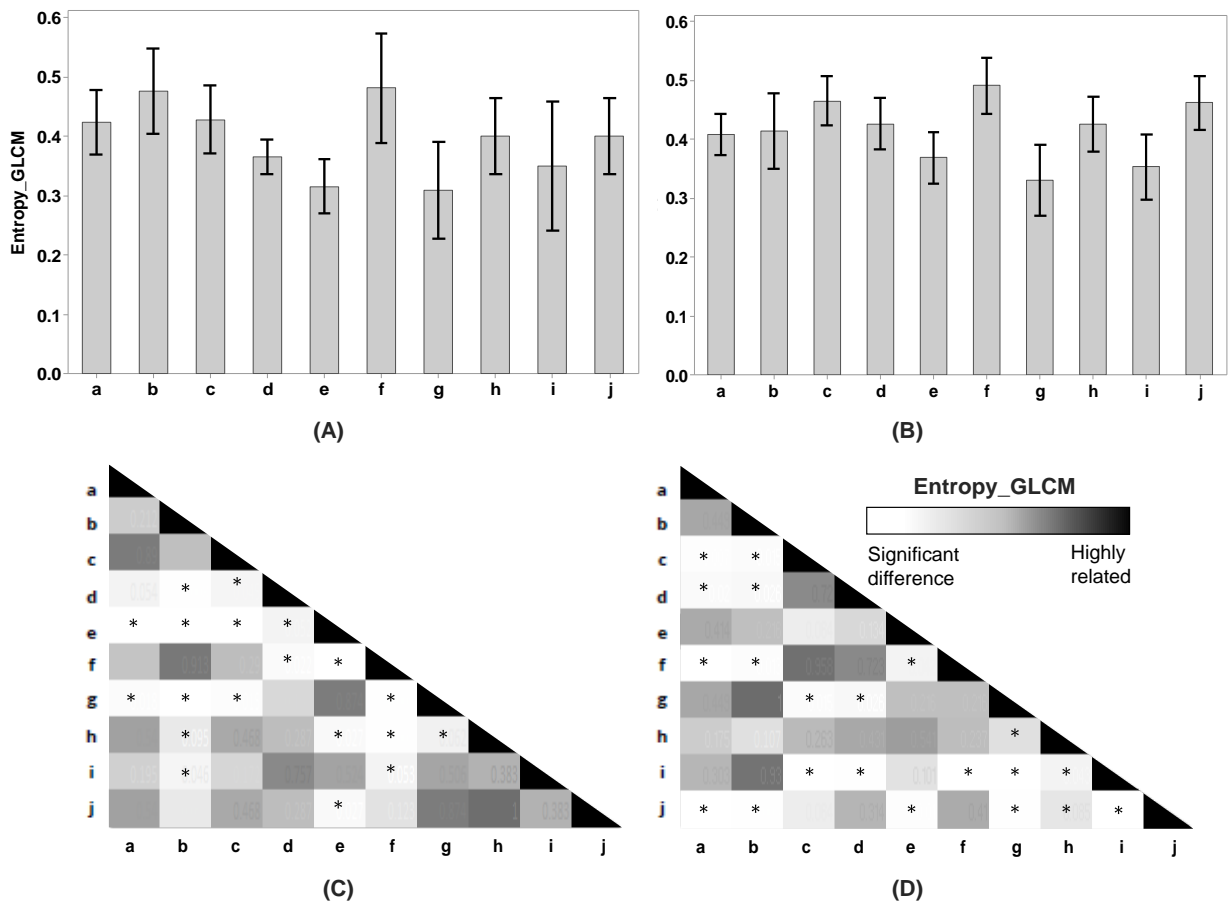

Figure Supplementary S8. GLCM texture analysis, entropy at 45º, results and comparison for (A) epidermis, and (B) dermis of ten body sites; Calculated gray-scale coded map of *p*-values between each pairs of body site (*p*-value < 0.05 considered as an acceptable significant difference indicated with *) for (C) epidermis, and (D) dermis. The letters from a-j demonstrate the following; (a) tip of nose, (b) preauricular, (c) volar forearm, (d) neck, (e) palm, (f) back, (g) thumb, (h) dorsal forearm, (i) sole, (j) calf.

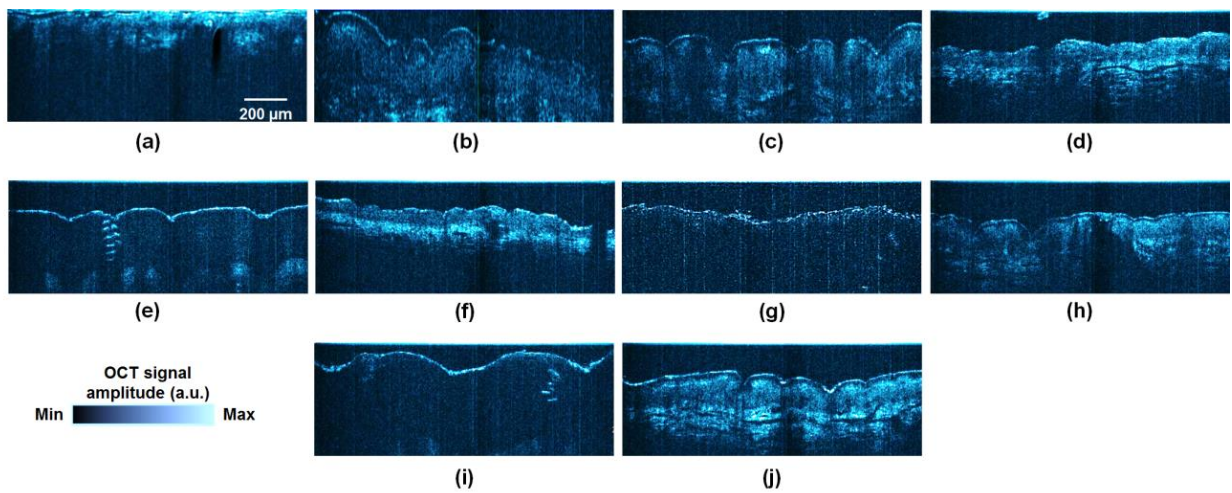

Figure Supplementary S9. HD OCT images of different sites of body including (a) tip of nose, (b) preauricular, (c) volar forearm, (d) neck, (e) palm, (f) back, (g) thumb, (h) dorsal forearm, (i) sole, (j) calf.

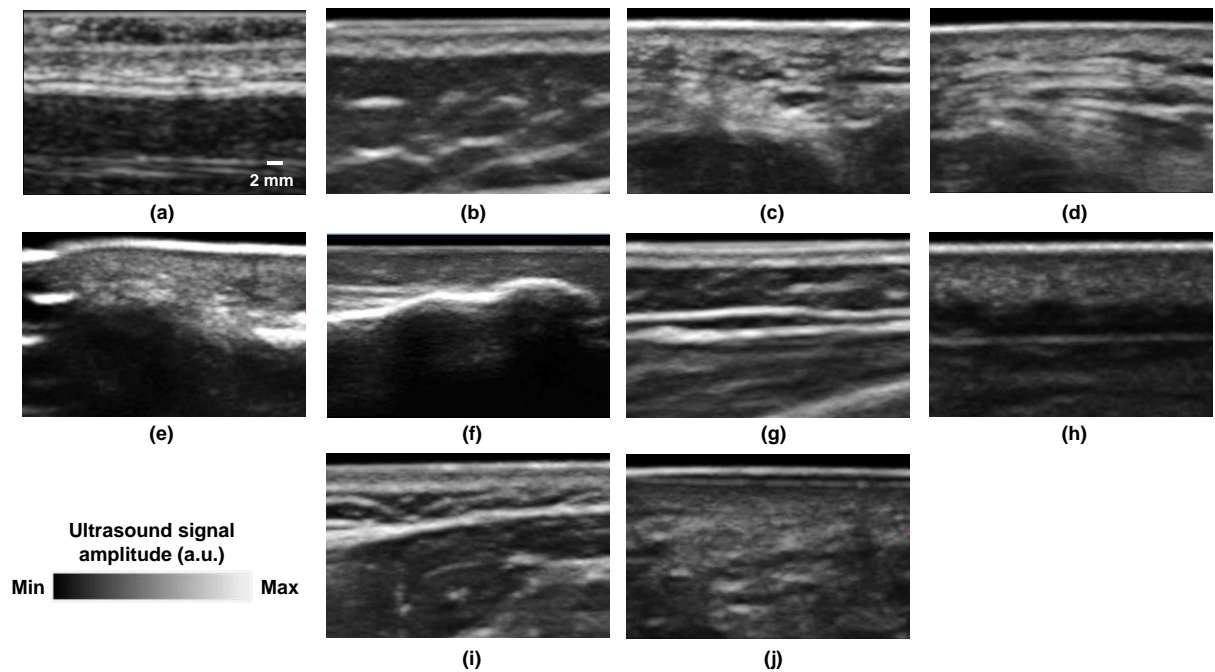

Figure Supplementary S10. Images obtained with a 15 MHz clinical ultrasound probe from different sites of body including (a) tip of nose, (b) preauricular, (c) volar forearm, (d) neck, (e) palm, (f) back, (g) thumb, (h) dorsal forearm, (i) sole, (j) calf.

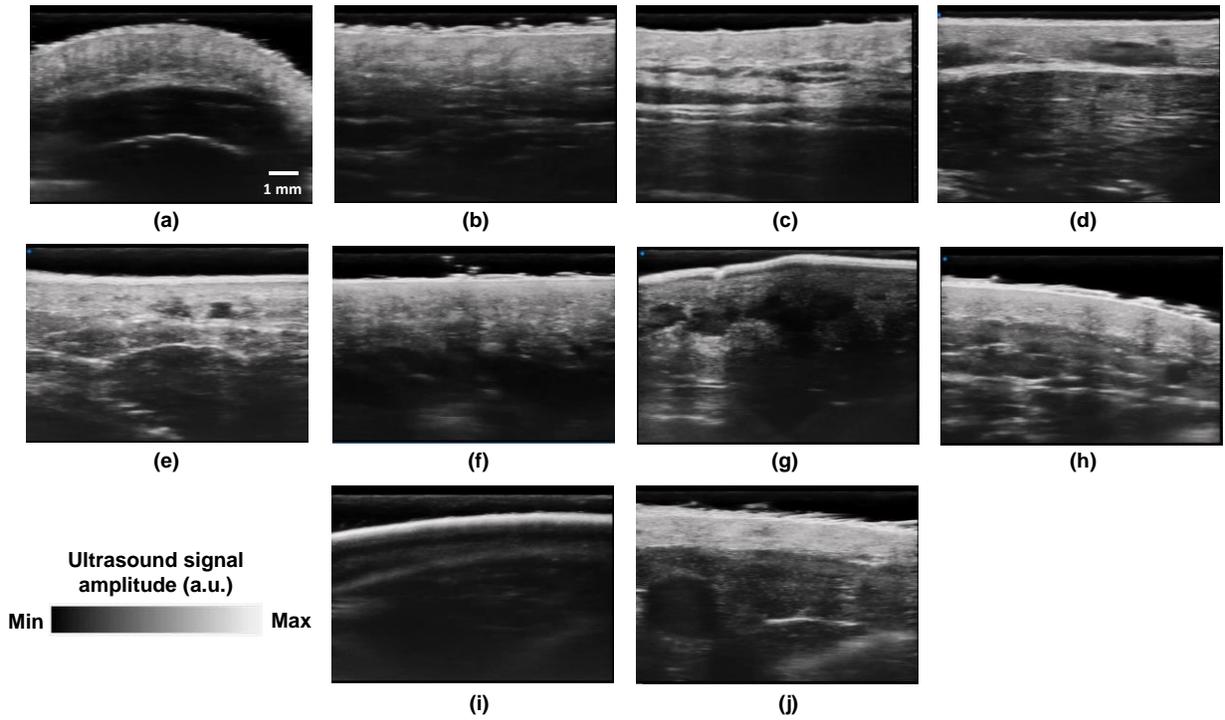

Figure Supplementary S11. Images obtained with a high frequency (VevoMD, 48 MHz) clinical ultrasound probe from different sites of body including (a) tip of nose, (b) preauricular, (c) volar forearm, (d) neck, (e) palm, (f) back, (g) thumb, (h) dorsal forearm, (i) sole, (j) calf.

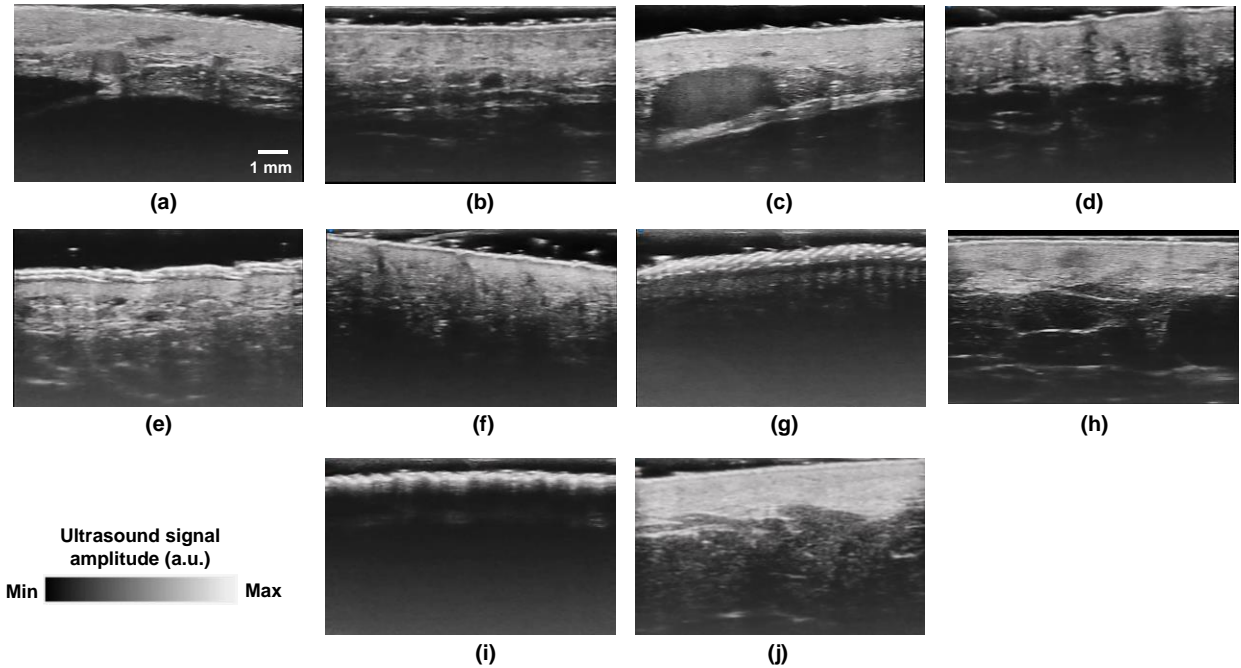

Figure Supplementary S12. Images obtained with a high frequency (VevoMD, 70 MHz) clinical ultrasound probe from different sites of body including (a) tip of nose, (b) preauricular, (c) volar forearm, (d) neck, (e) palm, (f) back, (g) thumb, (h) dorsal forearm, (i) sole, (j) calf.

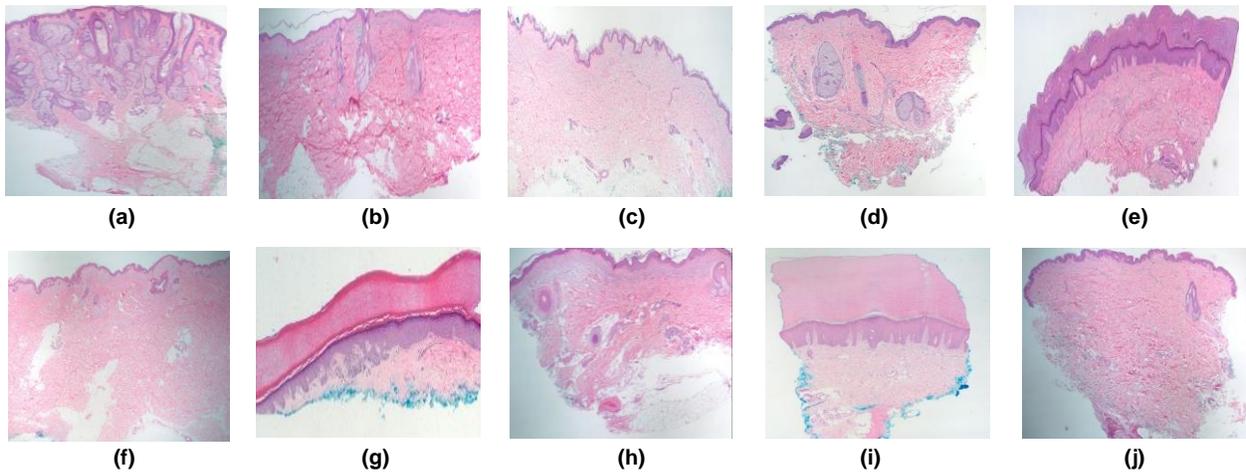

Figure Supplementary S13: histology images of healthy skin at different sites of body including (a) tip of nose, (b) preauricular, (c) volar forearm, (d) neck, (e) palm, (f) back, (g) thumb, (h) dorsal forearm, (i) sole, (j) calf.

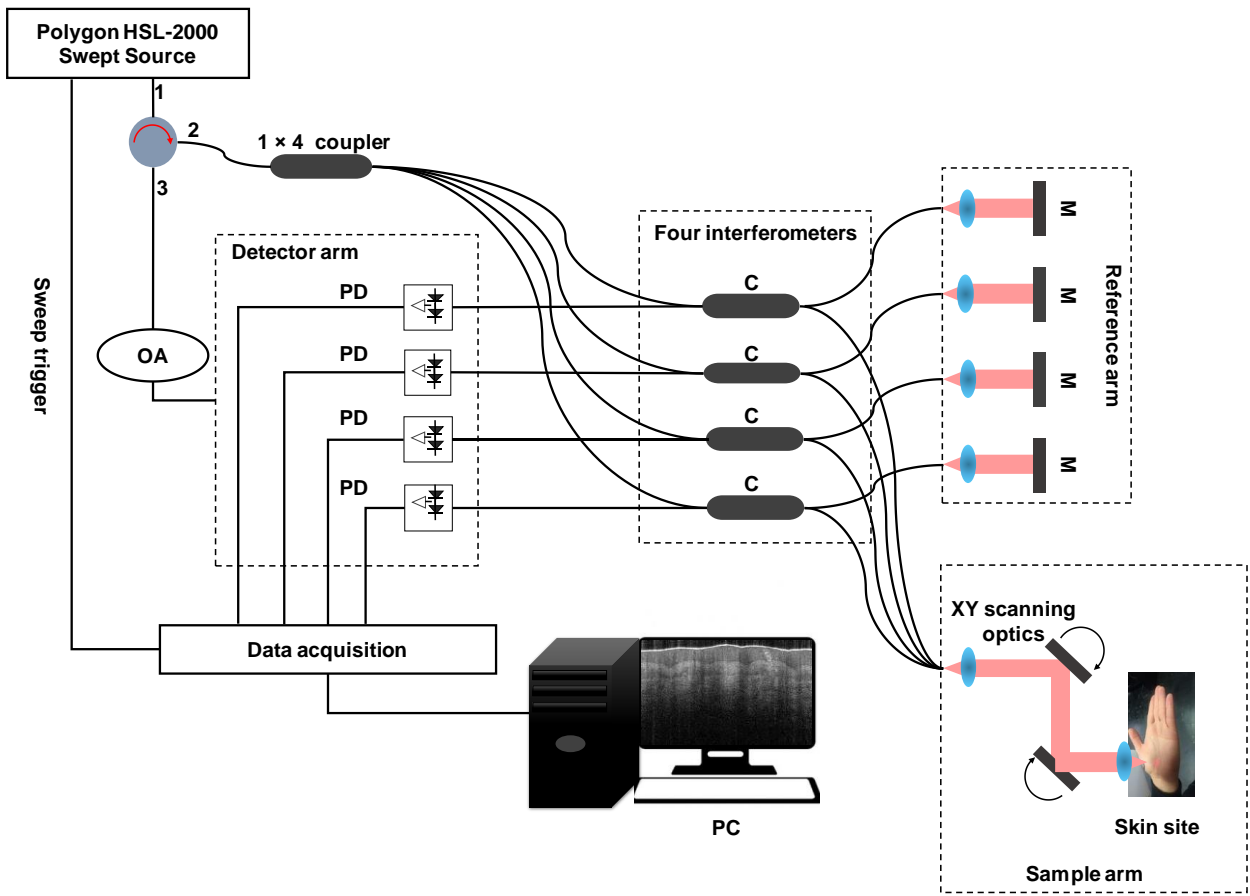

Figure Supplementary S14: Schematic diagram of the multi-beam swept source OCT; M: mirror, C: coupler, PD: photodetector, OA: optical attenuator, photograph of a hand is taken by S.A

9